\documentclass[
    reprint,
    twocolumn,
    showpacs,
    superscriptaddress,
    longbibliography,
    floatfix,
    aps,
    prl]{revtex4-2}

\usepackage{header}
\usepackage{comment}

\usepackage{xspace} 

\makeatletter
\begin{document}
\title{Creating squeezed and non-classical collective motional many-body states through stroboscopic Rydberg dressing}
\author{Roman Wu\ss ler}
\affiliation{Institut f\"ur Theoretische Physik and Center for Integrated Quantum Science and Technology, Universit\"at T\"ubingen, Auf der Morgenstelle 14, 72076 T\"ubingen, Germany}
\author{Chris Nill}
\affiliation{Institut f\"ur Theoretische Physik and Center for Integrated Quantum Science and Technology, Universit\"at T\"ubingen, Auf der Morgenstelle 14, 72076 T\"ubingen, Germany}
\affiliation{Institute for Applied Physics, University of Bonn, Wegelerstraße 8, 53115 Bonn, Germany}
\author{Sylvain de L\'es\'eleuc}
\affiliation{RIKEN Center for Quantum Computing (RQC), 351-0198 Wako, Japan}
\author{Christian Groß}
\affiliation{Physikalisches Institut and Center for Integrated Quantum Science and Technology, Universit\"{a}t T\"{u}bingen, Auf der Morgenstelle 14, 72076 T\"{u}bingen, Germany}
\author{Igor Lesanovsky}
\affiliation{Institut f\"ur Theoretische Physik and Center for Integrated Quantum Science and Technology, Universit\"at T\"ubingen, Auf der Morgenstelle 14, 72076 T\"ubingen, Germany}
\affiliation{School of Physics and Astronomy and Centre for the Mathematics and Theoretical Physics of Quantum Non-Equilibrium Systems, The University of Nottingham, Nottingham, NG7 2RD, United Kingdom}

\begin{abstract}
Realizing conditional quantum operations, e.g., quantum gates, for quantum computing and simulation requires controlled interactions between particles. Often, these interactions depend on the interparticle distance, and accordingly, an uncertainty of the relative particle position may translate into gate infidelities. We consider here a quantum computing platform based on an array of neutral atoms and present a method that allows to reduce the uncertainty of all interatomic distances. Our approach exploits the coupling between atomic motion and stroboscopically excited atomic Rydberg states. It allows to collectively squeeze the modes corresponding to interatomic displacements, thereby reducing distance fluctuations down to a fraction of the motional vacuum state. Furthermore, the method permits the creation of non-classical states with substantial Wigner negativity. These correlated states may allow reducing motional decoherence, increasing gate fidelity, and potentially yield a resource for quantum-enhanced metrology.
\end{abstract}

\maketitle
\emph{Introduction} --- Programmable arrays of laser-cooled atoms~\cite{Browaeys2020,Saffman2010, Adams2019,Bluvstein2023,Endres2016,Grotti2025}, polar molecules~\cite{Carr2009, Ruttley2025, Xiang2024}, and trapped ions~\cite{Eschner2003, Bruzewicz2019,Chernyshev2026} have emerged as leading platforms for quantum simulation, computation, and precision sensing. In all these systems, quantum operations are mediated by interactions that depend on interparticle distances~\cite{Monroe2021, Shao2024, Weber2017}. Thus, the coupling of electronic (spin) degrees of freedom to atomic motion is an inherent feature of any interatomic interaction-based quantum operation~\cite{Anderlini2007,Muller2014,Goerz2011}.

Controlling the motional state is therefore essential. This is typically achieved by optical trapping~\cite{Ahlheit2025,Grimm2000,Topcu2016} and laser cooling techniques~\cite{Kaufman2012, Clements2026} that suppress free atomic motion and thermal fluctuations to prepare atoms near their motional ground state. However, a fundamental residual quantum uncertainty persists even at zero temperature~\cite{Schine2022,Shaw2025a,Taylor2013}. Through the distance dependence of interactions, these quantum fluctuations translate directly into gate infidelities and decoherence~\cite{Robicheaux2021,Shi2020, Jenkins2012,Liu2021a,soto-garcia2026}.

A natural strategy to push below this ground-state limit is to prepare motionally squeezed states~\cite{Burd2024, Lo2015, Salvi2018,Rosiek2024}. Single-atom motional squeezing has recently been demonstrated experimentally~\cite{Lienhard2025}, which directly lowers the uncertainty of interatomic distances. However, in a many-atom array, an even stronger reduction can be achieved by synchronizing the motion of neighboring atoms. This requires squeezing the collective vibrational modes of the array, which is an inherent many-body operation.

Here, we demonstrate how many-body squeezed motional states can be realized in a one-dimensional chain of trapped atoms via stroboscopic Rydberg dressing~\cite{Koyluoglu2024, Feldmeier2024,Nill2025,Bharti2024}. A Bloch-Messiah decomposition~\cite{cariolaroReexaminationBlochMessiahReduction2016} of the stroboscopic time-evolution operator yields closed-form expressions for the motional state of arbitrarily long chains. For experimentally feasible parameters, we demonstrate that the variance, averaged over all nearest-neighbor distances, can be reduced down to 19\% of the ground-state value. Moreover, we show that by exploiting the non-quadratic nature of the Rydberg interaction potential, the generation of non-classical motional states~\cite{Rosiek2024,Bharti2024,Brown2023,Romero-Isart2011,Leibfried1996} with a Wigner negativity of 0.109 is possible.

\begin{figure*}
    \centering
    \includegraphics[width=\linewidth]{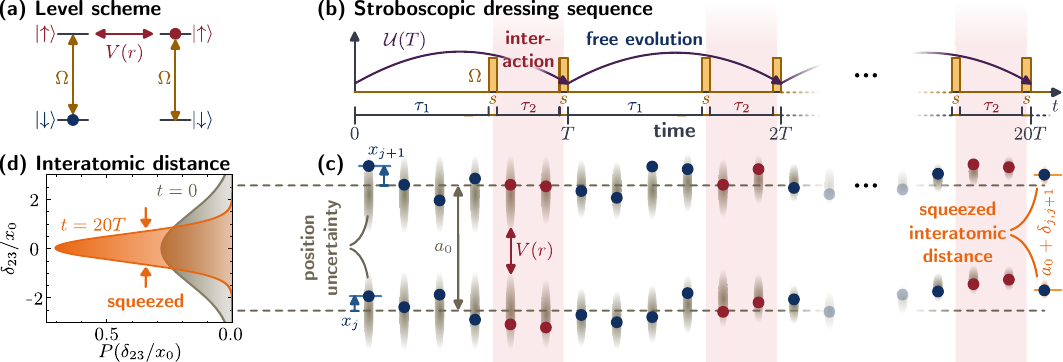}
\caption{
    \textbf{Stroboscopic Rydberg dressing for generating squeezed motional states.}
    (a)~Level scheme of the atom's electronic states: ground state $\ket{\downarrow}$ (blue) and Rydberg state $\ket{\uparrow}$ (red), coupled by a laser with Rabi frequency~$\Omega$. Neighboring atoms simultaneously excited to $\ket{\uparrow}$ interact via the van der Waals potential $V(r)$.
    (b)~The stroboscopic dressing protocol proceeds in periodic cycles of duration~$T$, repeated up to $20T$ as shown. Each cycle consists of free evolution of duration~$\tau_1$, a resonant $\pi$-pulse of duration~$s$ and Rabi frequency~$\Omega$ that transfers all atoms to~$\ket{\uparrow}$, an interaction interval of duration~$\tau_2$ during which atoms in $\ket{\uparrow}$ interact via $V(r)$ and develop motional correlations, and a second $\pi$-pulse returning all atoms to~$\ket{\downarrow}$.
    (c)~Schematic time evolution of two neighboring atoms confined in optical traps with equilibrium spacing~$a_0$ and displacements~$x_j$ from their respective equilibrium positions. Grey shading indicates the single-atom position variance of each atom. After multiple dressing cycles, the atoms' motions become correlated, reducing the fluctuations of the interatomic distance~$\delta_{j,j+1} = x_{j+1} - x_j$ below the ground-state level (squeezed interatomic distance, right).
    (d)~Probability distribution of the interatomic distance~$\delta_{23}$ at $t = 0$ (motional ground state, broad) and after $m = 20$ dressing cycles (narrow), demonstrating the reduction of distance fluctuations. Results are calculated from Eqs.~(\ref{eq:var_delta}) and~(\ref{eq:exp_delta}) for $N = 5$ atoms using the parameters of Eq.~(\ref{eq:parameter-set-1}).}
    \label{fig:cartoon}
\end{figure*}
\emph{Model} --- We consider a chain of $N$ two-level atoms with mass $m_0$ and mean spacing $a_0$. The internal states are an electronic ground state $\ket{\downarrow}$ and a Rydberg state $\ket{\uparrow}$ [Fig.~\ref{fig:cartoon}(a)]. We consider that both electronic states are trapped with the same trapping frequency $\omega$~\cite{Ahlheit2025,jansohn2025,Barredo2020}. When two neighboring atoms are simultaneously excited to the state $\ket{\uparrow}$, they interact via the van der Waals potential $V(r)=C_6/r^6$. This is a mechanism that couples electronic and motional degrees of freedom.
In the following, we restrict ourselves to the consideration of nearest-neighbor interactions~\cite{Browaeys2020,Saffman2010}. This is justified by the rapid $1/r^6$ decay of the van der Waals potential, which suppresses next-nearest-neighbor contributions by a factor of $(a_0/2a_0)^6=1/64$. The Hamiltonian reads ($\hbar=1$)
\begin{align}
    H_0=\omega\sum_{j=1}^Na_j^\dagger a_j+\sum_{j=1}^{N-1}V(a_0+x_{j+1}-x_j) n_j n_{j+1}.
    \label{eq:H_0.}
\end{align}
Here $n_j=\ketbra{\uparrow}{\uparrow}_j$ is the projector onto the Rydberg state of the $j$-th atom, and each trapping potential is modeled as a one-dimensional harmonic oscillator with ladder operators $a_j$ and $a_j^\dagger$, describing the atomic motion along the chain axis. The displacement of atom $j$ from its equilibrium position is $x_j=x_0(a_j+a_j^\dagger)$, with the harmonic oscillator length $x_0=1/\sqrt{2m_0\omega}$.

We laser-excite the atoms with a stroboscopic pulse sequence, shown in Fig.~\ref{fig:cartoon}(b). It starts from the product 
ground state $\ket{\Psi(0)}=\ket{\downarrow,0}^{\otimes N}$ and proceeds in repeated cycles of 
duration $T$. The time evolution operator for $m$ cycles is given by
\begin{align}
    \mathcal{U}(mT)=\Bigl(\underbrace{e^{-is(H_\mathrm{L}+H_0)}}_{\mathcal U_\pi}\, 
    e^{-i\tau_2H_0}\, \mathcal U_\pi\, e^{-i\tau_1H_0}\Bigr)^m,
\end{align}
and to be read from right to left: each cycle starts with free evolution under the Hamiltonian $H_0$ [Eq.~(\ref{eq:H_0.})] for time $\tau_1$, followed by a resonant 
$\pi$-pulse $\mathcal{U}_\pi$ of duration $s$ that---assuming it is sufficiently strong---simultaneously excites all atoms to $\ket{\uparrow}$. This short pulse is followed by an interval of length $\tau_2$ in which the atoms are in the Rydberg state and the evolution is governed by $H_0$. Finally, a second $\pi$-pulse 
returns all atoms to $\ket{\downarrow}$. The laser Hamiltonian
\begin{align}
    H_\mathrm{L}=\frac{\pi}{2s}\sum_{j=1}^N\Bigl(\ketbra{\downarrow}{\uparrow}_j
    e^{-i\kappa x_j}+\ketbra{\uparrow}{\downarrow}_je^{i\kappa x_j}\Bigr)
\end{align}
generates the $\pi$-pulses. Here, $\kappa$ is the laser wave vector projected along the 
chain axis, and the Lamb-Dicke parameter $\eta=\kappa x_0$ quantifies the 
photon-recoil-induced motional excitation~\cite{Robicheaux2021,Parvej2025,Apolin2026}. We consider the Rydberg interaction interval $\tau_2$ to be orders of 
magnitude shorter than the typical Rydberg state lifetime of $\sim$$\SI{50}{\micro\second}$, 
such that spontaneous emission during each dressing cycle is negligible~\cite{Cohen2021, Picken2018,Schine2022}.

\emph{Analytical treatment} --- To enable an analytical treatment, we work in the limit of fast and strong $\pi$-pulses, $\Omega=\pi/s\gg\omega,V(r)$, in which $H_0$ can be neglected during the pulses. This simplifies the operator describing the $\pi$-pulses to $\mathcal{U}_\pi=e^{-isH_\mathrm{L}}$. Within this limit, the application of $\mathcal{U}_\pi$ immediately switches all atoms between the electronic ground state and the Rydberg state. This approximation guarantees that all atoms remain in the electronic ground state at integer multiples of $T$. Its validity---quantified by the residual Rydberg population at stroboscopic times---improves with decreasing $s$, as we validate in the End Matter. The time-evolution operator $\mathcal{U}(mT)$ therefore acts solely on the motional degrees of freedom.

During the Rydberg interaction phase $\tau_2$, all atoms simultaneously occupy the Rydberg state and experience mutual repulsion, which couples the motion of neighboring atoms, [Fig.~\ref{fig:cartoon}(c)]. This coupling transforms the motional eigenmodes of the chain from local single-atom vibrations into collective normal modes, which can be written as $b_k = \sum_j Q_{kj} a_j$, i.e., linear combinations of the single-atom modes $a_j$ with transformation matrix elements $Q_{1j}=1/\sqrt{N}$ and $Q_{kj}= \sqrt{2/N}\cos\{{(j-1)(k-\frac{1}{2})\pi}/{N}\}$ for $2\le j\le N$. While $b_1$ describes the center-of-mass motion of the entire chain, all higher modes $b_k$ with $k\geq2$ capture relative motion between atoms and thus directly govern the interatomic distances. To obtain closed-form expressions for the dynamics of these modes, we expand the van der Waals potential to second order around the equilibrium distance $a_0$, yielding $V(r)\approx V_0 + V_1\,(r-a_0) + \tfrac{1}{2}V_2\,(r-a_0)^2$, where $V_0=V(a_0)$, $V_1=\partial V/\partial r|_{a_0}$, and $V_2=\partial^2 V/\partial r^2|_{a_0}$ are the value, gradient, and curvature of the potential at equilibrium. This approximation is controlled by $x_0/a_0$; higher-order corrections are suppressed by successive powers of $x_0/a_0$ and are explored later on.
Within this quadratic approximation, a Bloch-Messiah decomposition of $\mathcal{U}(mT)$ in the basis of the normal modes $b_k$ (see Supplemental Material~\cite{Supplement}) yields a physically transparent structure: the protocol independently squeezes, displaces, and rotates each normal mode, and, up to a global phase, the time evolution operator can be written as
\begin{align}
    \mathcal{U}(mT)=\prod_{k=1}^N\mathcal{S}_k(Z_k)\mathcal{D}_k(J_k)\mathcal{R}_k(\beta_k).
    \label{eq:time_evolution_operator}
\end{align}
Here, $\mathcal{S}_k(Z_k)=\exp\big\{\frac{Z_k^*}{2}b_k^2-\frac{Z_k}{2}(b_k^\dagger)^2\big\}$, $\mathcal{D}_k(J_k)=\exp\{J_kb_k^\dagger-J_k^*b_k\}$, and $\mathcal{R}_k(\beta_k)=\exp\{i\beta_kb_k^\dagger b_k\}$ are the single-mode squeezing, displacement, and rotation operators. The vibrational modes hence undergo Gaussian dynamics, and all parameters $Z_k$, $J_k$, $\beta_k$ are known in closed form, see End Matter. This result holds for chains of arbitrary length $N$.

The displacement $\mathcal D_k(J_k)$ of relative modes $k\geq2$ depends on the potential gradient~$V_1$, which imprints a force between atoms.
The center-of-mass displacement $J_1$ is independent of $V_1$. It arises solely due to momentum transfer of the photon recoil during the $\pi$-pulses, quantified by the Lamb-Dicke parameter $\eta$.

Squeezing $\mathcal S_k(Z_k)$ of relative modes [Fig.~\ref{fig:cartoon}(d)] originates from the interaction potential curvature $V_2$, which modifies the relative motional mode trap frequencies. During stroboscopic dressing, the system alternates between free evolution (duration $\tau_1$) at the bare frequency $\omega$ and an interaction phase (duration $\tau_2$). This interaction shifts each relative mode to an effective frequency $\tilde{\omega}_k = \sqrt{\omega^2 + 2\lambda_k x_0^2 V_2 \omega}$, where the geometric factor $\lambda_k = 4\sin^2\{{(k-1)\pi}/2N\}$ encodes the mode index and chain length.
This periodic alternation between frequencies acts as a parametric modulation that squeezes every relative mode ($Z_k \neq 0$); if $V_2 = 0$, squeezing vanishes. Analogous to single-atom squeezing via trap frequency modulation~\cite{Lienhard2025}, this mechanism drives collective normal modes and creates correlations between atoms.

\begin{figure}
    \centering
    \includegraphics[width=\linewidth]{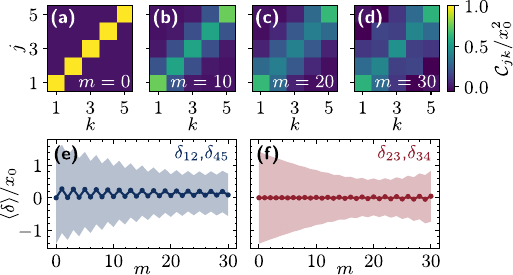}
    \caption{%
    \textbf{Build-up of positional correlations and dynamics of interatomic distances.}
    (a)~-~(d)~Positional covariance matrix $\mathcal C_{jk}/x_0^2$ after $m=0,10,20$ and 30 dressing cycles. Starting from an uncorrelated motional ground state (a), the protocol induces positional correlations between neighboring atoms (b), (c). At $m=20$~(c), reduced diagonal elements (single-atom variances) and increased off-diagonal elements (inter-atom correlations) combine to yield significant squeezing of the interatomic distances. As the sequence continues~(d), variances grow again while correlations persist. (e),~(f)~Mean interatomic distances $\langle\delta\rangle/x_0$ (solid lines) in a chain of $N=5$ atoms, with shading indicating one standard deviation about the mean versus dressing cycle $m$ for (e)~outer atom pairs ($\delta_{12}$, $\delta_{45}$) and (f)~inner atom pairs ($\delta_{23}$, $\delta_{34}$). Outer atoms oscillate with a larger amplitude due to asymmetric boundary forces, while inner atoms remain closer to their equilibrium position.
    The choice $\omega T\gtrsim\pi$ keeps the mean displacement bounded, see Supplemental Material~\cite{Supplement}. In all panels, we use Eqs.~(\ref{eq:var_delta}),~(\ref{eq:exp_delta}), and the parameters from Eq.~(\ref{eq:parameter-set-1}), which were optimized to minimize the variance of the interatomic distances after $m=20$ dressing cycles.}
    \label{fig:mean_and_covariances}
\end{figure}

\emph{Many-body motional squeezing} --- Having a closed-form expression for the time evolution of the motional state, our goal is now to engineer the stroboscopic dressing such that quantum fluctuations of the interatomic distances $\delta_{j,j+1}=x_{j+1}-x_j$ are reduced below their ground-state value [Fig.~\ref{fig:cartoon}(d)]. To quantify these fluctuations we use the variance
\begin{align}
        \mathrm{Var}(\delta_{j,j+1})=\mathrm{Var}(x_{j+1})+\mathrm{Var}(x_j)-2\,\mathcal C_{j,j+1}.
        \label{eq:var_decomp}
\end{align}
Here, $\mathcal{C}_{j,k}=\langle x_jx_k\rangle-\langle x_j\rangle\langle x_k\rangle$ is the covariance matrix which describes correlations between atoms. The diagonal elements $\mathrm{Var}(x_j)=\mathcal{C}_{j,j}$ represent the individual variances. Equation~(\ref{eq:var_decomp}) shows two complementary mechanisms to reduce interatomic distance uncertainty: (i)~squeezing the individual atomic positions to minimize $\mathrm{Var}(x_j)$, and (ii)~correlating the motion of neighboring atoms to achieve $\mathcal{C}_{j,j+1}>0$, meaning that the atoms' motion synchronizes. Single-atom squeezing, which was recently demonstrated in~\cite{Lienhard2025}, exploits solely the first mechanism.
We show now that the pulsed excitation of Rydberg states additionally synchronizes the atomic positions, exploiting both mechanisms simultaneously. The build-up of positive motional correlations constitutes a many-body effect, enabling a further suppression of distance uncertainty beyond what single-atom squeezing could achieve.

Figure~\ref{fig:mean_and_covariances} illustrates the build-up of positional correlations and the dynamics of interatomic distances for a chain with $N=5$ atoms. Starting from the motional ground state, which exhibits no interatomic correlations~(a), stroboscopic dressing progressively generates positive off-diagonal entries in the covariance matrix~$\mathcal{C}_{jk}$~(b),(c) while simultaneously reducing the diagonal entries (single-atom variances). For the chosen parameters from Eq.~(\ref{eq:parameter-set-1}), after $m=20$ cycles~(c), reduced diagonal elements and increased off-diagonal elements combine to yield significant squeezing of the interatomic distances. Here, the variance, averaged over all nearest-neighbor distances, is approximately 19\% of the initial ground-state value [see Eq.~\eqref{eq:var_decomp}]. Notably, the build-up of interatomic correlations accounts for a substantial portion of this effect: the covariance term $-2\mathcal{C}_{j,j+1}$ alone lowers the averaged variance by an additional 28\% of the ground-state value, highlighting the significant advantage of establishing interatomic correlations over single-atom squeezing~\cite{Lienhard2025}. As the sequence continues, the single-atom variances grow again while the positional correlations persist~(d). In panels (e) and (f), outer atom pairs ($\delta_{12}$, $\delta_{45}$) exhibit larger mean displacements than inner atom pairs ($\delta_{23}$, $\delta_{34}$), since inner atoms experience counteracting nearest-neighbor forces from both sides.

\emph{Non-classical states}\label{sec:nonclassical} --- The analytical results derived via the Bloch-Messiah decomposition in the previous discussion rely on the expansion of the interaction potential $V(r)$ around the equilibrium interatomic distances. In the following, we demonstrate that higher-order terms break the Gaussianity of the evolution and how this can be leveraged to generate non-classical states. As previously noted, the quadratic approximation is valid when $x_0/a_0\ll1$.
To explore the boundaries of our analytical model, we compare the system's dynamics under the approximated quadratic potential with those under the exact van der Waals potential. As an indicator, we use the Wigner function $W(\delta_{12}, v_{12})$ of the relative coordinate $\delta_{12}$ and relative velocity $v_{12}=v_2-v_1$ for $N=2$ atoms after $m=2$ dressing cycles. The single-atom velocities are represented by $v_j=iv_0(a_j^\dagger-a_j)$, where $v_0=\omega x_0$ denotes the characteristic ground-state velocity.

\begin{figure}
    \centering
    \includegraphics[width=\linewidth]{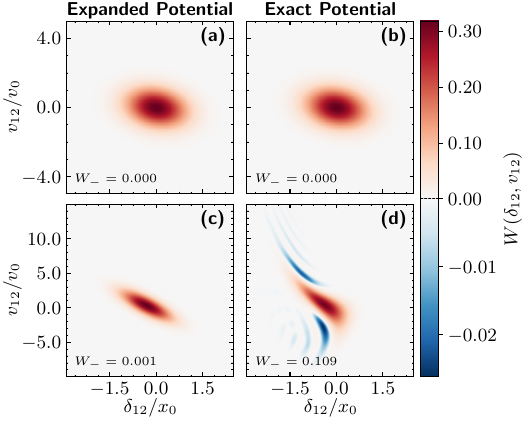}
    \caption{
    \textbf{Non-classical motional states.}
    Wigner functions $W(\delta_{12},v_{12})$ of the relative coordinate $\delta_{12}$ and relative velocity $v_{12}$ for $N=2$ atoms after $m=2$ dressing cycles, computed using the second-order Taylor expansion of the interaction potential (left column) and the exact van der Waals potential (right column).
    (a),~(b)~For parameters from Eq.~(\ref{eq:parameter-set-1}) with $x_0/a_0\approx0.009$, the harmonic approximation faithfully reproduces the exact result: both Wigner functions are Gaussian and squeezed. (c),~(d) For parameters from Eq.~(\ref{eq:parameter-set-2}) with $x_0/a_0\approx0.057$, the harmonic approximation yields a Gaussian squeezed state~(c), but the exact dynamics produces a strongly non-Gaussian distribution with pronounced negative regions, indicated with blue in~(d), signaling genuine quantum non-classicality induced by higher-order terms of the van der Waals potential.
    }
    \label{fig:wigner}
\end{figure}
For $x_0/a_0\approx0.009$ [Eq.~(\ref{eq:parameter-set-1})], the atomic wave function is well localized compared to $a_0$, and the quadratic approximation is excellent. As seen in Fig.~\ref{fig:wigner}, the incorporation of higher order terms does not alter the result, making the Wigner functions in ~(a) and ~(b) indistinguishable, both showing the squeezed state predicted by the Bloch-Messiah decomposition. This confirms that the analytical framework of the previous section faithfully captures the full dynamics in this regime. The situation changes qualitatively for $x_0/a_0\approx0.057$ [Eq.~(\ref{eq:parameter-set-2})], where the wave function is so wide that it is also sensitive to changes in the potential's curvature. While the quadratic model still predicts a Gaussian squeezed state~(c), the exact dynamics generates a Wigner function with pronounced negative regions~(d), reaching a Wigner negativity of $W_-=0.109$. This is an unambiguous signature of quantum non-classicality~\cite{kenfackNegativityWignerFunction2004}, which reveals that the higher-order contributions of the van der Waals interaction, often treated as a mere correction, act as a resource for generating non-classical motional states.

\emph{Discussion and outlook} --- The protocol is well within reach of current experiments. Both parameter sets [Eqs.~(\ref{eq:parameter-set-1}) and~(\ref{eq:parameter-set-2})] employ $^{87}$Rb atoms in optical tweezer arrays. Established sideband-cooling techniques prepare the required initial state $\ket{\downarrow,0}^{\otimes N}$, with near-unity ground-state fidelity recently demonstrated~\cite{Shaw2025a}.

A fundamental requirement of our analytical framework is that the laser drive constitutes by far the largest energy scale in the system ($\Omega\gg\omega, V(r)$). With Rabi frequencies of several hundreds of megahertz, this is satisfied for both configurations. Furthermore, these driving frequencies remain smaller than the energy separation to adjacent Rydberg levels, which we elaborate in the End Matter.

Parameter set Eq.~(\ref{eq:parameter-set-1}) assumes a spacing of $a_0=\SI{2.7}{\micro\meter}$ and a trap frequency $\omega=2\pi\times\SI{100}{\kilo\hertz}$ — values achievable in optical tweezer arrays~\cite{Lienhard2025}. The non-classical regime [Eq.~(\ref{eq:parameter-set-2})] requires a smaller interatomic spacing $a_0=\SI{1}{\micro\meter}$, corresponding to tighter traps already accessible with state-of-the-art setups~\cite{Bharti2024}. For these configurations, twenty dressing cycles are completed in $\sim\SI{100}{\micro\second}$ ($T=\SI{4.98}{\micro\second}$) and $\sim\SI{600}{\micro\second}$ ($T=\SI{30}{\micro\second}$), respectively — well within atomic coherence times of several milliseconds.

Furthermore, the impact of anharmonic tweezer potentials is analyzed in the Supplemental Material~\cite{Supplement}: for current standard optical tweezers, the Wigner function experiences shearing and non-Gaussian deformations. While realizing purely Gaussian, strongly squeezed states will require improved trap harmonicity, this inherent anharmonicity is not strictly a limitation. Instead, it can be actively leveraged as a resource for generating genuine non-classical many-body motional states~\cite{Grochowski2025}.

Looking ahead, our findings suggest several immediate technological applications:

(i)~Motional decoherence currently limits the fidelity of distance-sensitive operations, such as Rydberg interaction gates~\cite{Saffman2010,Robicheaux2021,Tsai2025}. Integrating the squeezing protocol could improve gate performance by mitigating positional uncertainty during the gate operation. 

(ii)~Our protocol could directly enhance the performance of cooperative atom-light interfaces that exhibit sub- and superradiance~\cite{Rui2020,Olmos2025a,Srakaew2023}. In these systems, positional uncertainty worsens the delicate interference mechanisms necessary for collective emission, leading to a broadened linewidth. Utilizing the proposed squeezing approach to suppress interatomic distance fluctuations would mitigate this decoherence.

(iii)~The engineered reduction of distance fluctuations directly translates to a narrowed linewidth of distance-dependent spectral features. Squeezed motional states could thus act as a resource for high-precision measurements~\cite{Chalopin2018,Li2023a} of Rydberg interaction coefficients and fundamental atomic properties.

(iv)~The facilitated excitation of Rydberg atoms (anti-blockade) is rather sensitive to the interatomic distances of neighboring atoms~\cite{Magoni2024, Brady2025}. Therefore, the proposed motional squeezing could stabilize these dynamics.

\emph{Data Availability} --- Data and code corresponding to this manuscript are publicly available~\cite{zenodo}.

 \emph{Acknowledgments} --- 
This work received funding from the Horizon Europe program HORIZON-CL4-2022-QUANTUM-02-SGA via the project 101113690 (PASQuanS2.1), the Deutsche Forschungsgemeinschaft within the research units FOR 5413 (Grant No. 465199066) and FOR 5522 (Grant No. 499180199) as well as from the European Union through the ERC grant OPEN-2QS (Grant No. 101164443). We also acknowledge funding through JST-DFG 2024: Japanese-German Joint Call for Proposals on “Quantum Technologies” (Japan-JST-DFG-ASPIRE 2024) under JST Grand No. JPMJAP24C2 and DFG Grant No. 554561799, and from the Alfried Krupp von Bohlen and Halbach Foundation. RW acknowledges support from the German Academic Scholarship Foundation (Studienstiftung des deutschen Volkes).
SdL thanks V. Magro for discussion and critical reading of the manuscript. 

\bibliography{references}

\appendix
\setcounter{equation}{0}
\setcounter{figure}{0}
\setcounter{table}{0}
\makeatletter
\renewcommand{\theequation}{EM\arabic{equation}}
\renewcommand{\thefigure}{EM\arabic{figure}}
\section{End Matter}
\subsection{Experimental considerations}
For the investigation in the main text, we use two experimental parameter sets. The first is
\begin{equation}
    \begin{aligned}
    a_0&= \SI{2.7}{\micro\meter}, \, \omega = 2\pi \times \SI{100}{\kilo\hertz} \, (x_0\approx \SI{24.1}{\nano\meter}), \\ C_6& = \SI{4.11}{\giga\hertz\,\micro\meter^6}, \, \eta = 0.121, \, \tau_2 = \SI{890}{\nano\second}, \, \\
    T& = \SI{4.98}{\micro\second}, \, s = \SI{5}{\nano\second} \, (\Omega = \pi \times \SI{200}{\mega\hertz}).
    \label{eq:parameter-set-1}
    \end{aligned}
\end{equation}
Here, the van der Waals coefficient $C_6$ corresponds to $^{87}$Rb atoms excited to the state $\ket{45S}$, which has a lifetime of $\tau_l=\SI{48.6}{\micro\second}$ at $\SI{300}{\kelvin}$. The pairstate $\ket{45S, 45S}$ exhibits an energy separation of $D=\SI{3.5}{\giga\hertz}\approx5.6\Omega$ to the nearest relevant pairstate $\ket{44P, 45P}$. For this configuration, the ratio of oscillator length to interatomic separation is $x_0/a_0\approx0.009$.

The second parameter set is
\begin{equation}
    \begin{aligned}
        a_0&= \SI{1}{\micro\meter}, \, \omega = 2\pi \times \SI{18}{\kilo\hertz} \, (x_0\approx \SI{56.8}{\nano\meter}), \\ C_6& = \SI{87}{\mega\hertz\,\micro\meter^6}, \, \eta = 0.286, \, \tau_2 = \SI{20}{\nano\second}, \, \\
        T& = \SI{30}{\micro\second}, \, s = \SI{1}{\nano\second} \, (\Omega = \pi \times \SI{1}{\giga\hertz}).
        \label{eq:parameter-set-2}
    \end{aligned}
\end{equation}
In this case, $C_6$ corresponds to the state $\ket{35S}$ of $^{87}$Rb with a lifetime of $\tau_l=\SI{24.2}{\micro\second}$ at $\SI{300}{\kelvin}$. The energy separation of the pairstate $\ket{35S, 35S}$ to the next relevant pairstate $\ket{34P, 35P}$ is $D=\SI{9.15}{\giga\hertz}\approx2.9\Omega$ and $x_0/a_0=0.057$.

For both parameter sets, the Lamb-Dicke parameter $\eta$ is calculated assuming a two-photon excitation scheme with typical wavelengths of $\lambda_1=\SI{780}{\nano\meter}$ and $\lambda_2=\SI{480}{\nano\meter}$, resulting in an effective laser wave vector $\kappa=2\pi(1/\lambda_2-1/\lambda_1)$. The $C_6$ coefficients and all respective level spacings were computed using the pairinteraction software~\cite{Mogerle2026}.

To obtain an experimental protocol that maximizes squeezing, several aspects have to be taken into account.
As we show later, the achievable squeezing grows with the curvature $V_2$ of the interaction potential (see the squeezing parameter $Z_k$ in the End Matter further below) and the total interaction duration $m\tau_2$. The protocol hereby has to fulfill the following constraints:

(i)~\emph{Strong pulses:} For a given van der Waals coefficient $C_6 \sim V(r)$ the Rabi frequency has to be chosen to break the Rydberg blockade ($\Omega \gg V(r)$), such that all ground state atoms can be excited to a Rydberg state.

(ii)~\emph{Resolution of Rydberg Pair States:} The Rabi frequency $\Omega$ is strictly upper-bounded by the energy separation to adjacent Rydberg pair states $D \gg \Omega$. This prevents unwanted off-resonant excitations of these, justifying the approximation of strict two-level atoms.

(iii)~\emph{Total interaction duration:} While increasing the total interaction duration $m\tau_2$ enhances the achievable squeezing, it directly elevates the probability of spontaneous emission due to the finite lifetime $\tau_l$ of the Rydberg state. Therefore, the protocol requires $m\tau_2\ll \tau_l$.

To find an optimal tradeoff among these constraints, the principal quantum number of the Rydberg state can be carefully selected. Increasing it, increases both the lifetime $\tau_l$ and the van der Waals coefficient $C_6$, whereas the energy separation $D$ decreases.

Based on these principles, parameter sets [Eqs.~\eqref{eq:parameter-set-1} and \eqref{eq:parameter-set-2}] for Rubidium have been chosen. The validity of the fast pulse approximation for Eq.~\eqref{eq:parameter-set-1} is subsequently evaluated through a numerical study.

\subsection{Validity of the fast pulse approximation}
The analytical treatment in the main text relies on the fast-pulse approximation, in which the free Hamiltonian $H_0$ is neglected during the $\pi$-pulses, simplifying $\mathcal{U}_\pi = e^{-is(H_\mathrm{L}+H_0)} \approx e^{-isH_\mathrm{L}}$. A quantitative measure of the approximation quality is the residual Rydberg excitation $\langle n_\mathrm{Ryd} \rangle=\sum_j\langle n_j\rangle$ at stroboscopic times $t=mT$, which vanishes identically for perfect $\pi$-pulses but acquires a finite value when $H_0$ is retained during the pulse.

To assess this quantitatively for finite but small $s$, where the laser drive $\Omega=\pi/s\gg\omega, V(r)$ remains the dominant energy scale, we employ time-dependent perturbation theory to obtain 
\begin{align*}
    \mathcal{U}_\pi=&e^{-isH_\mathrm{L}}\biggl(1-is\int_0^1d\tau e^{i\tau sH_\mathrm{L}}H_0e^{-i\tau sH_\mathrm{L}}+\mathcal{O}(s^2)\biggr).
\end{align*}
The leading correction to $U_\pi$ is proportional to $sH_0$. Since the Rydberg interaction $V$ is by far the largest energy scale in the free Hamiltonian ($V\gg\omega$) for both parameter sets from Eqs.~(\ref{eq:parameter-set-1}) and~(\ref{eq:parameter-set-2}), we can extract it to define a dimensionless control parameter $sV$. Consequently, at any stroboscopic time $t=mT$, the state of the system acquires a correction that is linear in this parameter: $\ket{\Psi(mT)}=\ket{\Psi_\mathrm{ideal}(mT)}+sV\ket{\delta\Psi}+\mathcal{O}((sV)^2)$. Because the ideal stroboscopic pulse sequence returns all atoms perfectly to the electronic ground state, it follows that $n_j\ket{\Psi_\mathrm{ideal}(mT)}=0$. Therefore, the residual Rydberg population becomes $\langle n_\mathrm{Ryd}\rangle=(sV)^2\bra{\delta\Psi}\sum_jn_j\ket{\delta\Psi}+\mathcal{O}((sV)^3)$.

\begin{figure}
    \centering
    \includegraphics[width=\linewidth]{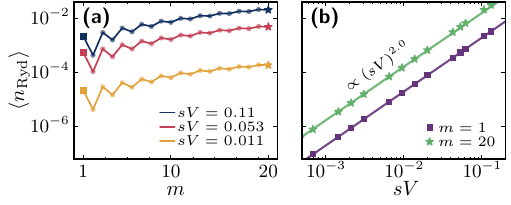}
    \caption{\textbf{Residual Rydberg excitation $\langle n_\mathrm{Ryd} \rangle$ at stroboscopic times for $N=2$ atoms.}
    (a) $\langle n_\mathrm{Ryd} \rangle$ versus cycle number $m$. The red curve ($sV=0.053$) represents the parameter set from Eq.~(\ref{eq:parameter-set-1}). For the other curves, the control parameter $sV$ was changed.
    (b) $\langle n_\mathrm{Ryd} \rangle$ as a function of $sV$ for $m = 1$ and $m=20$ [see markers in (a)]. The clean $\propto (sV)^{2}$ scaling confirms that the fast-pulse approximation is controlled by the control parameter $sV$. It breaks down when the Rydberg interaction energy $V$ becomes comparable to the Rabi frequency $\Omega=\pi/s$.}
    \label{fig:em-fast-pulses}
\end{figure}

Figure~\ref{fig:em-fast-pulses} validates this $(sV)^2$ scaling numerically. Panel~(a) shows $\langle n_\mathrm{Ryd} \rangle$ as a function of the cycle number $m$ for different values of the control parameter $sV$. While the accumulated error expectedly grows with the number of dressing cycles, it remains highly suppressed for small values of $sV$. Panel~(b) makes the underlying scaling explicit, plotting $\langle n_\mathrm{Ryd} \rangle$ directly against $sV$. The clean $(sV)^2$ collapse of the residual excitation for $m = 1$ and $m=20$ confirms that the fast-pulse approximation is well-controlled as long as $sV \ll 1$, which is equivalent to $\Omega\gg V$.

\subsection{Closed form expression of the time evolution operator}
Within the limits discussed before, the time evolution operator $\mathcal U(mT)$ can be expressed in closed form [Eq.~(\ref{eq:time_evolution_operator})]. An exhaustive derivation can be found in the Supplemental Material~\cite{Supplement}.
The parameters are
\begin{equation*}
    \begin{aligned}
        Z_1&=0,\\
        J_1&=\sqrt{N}i\eta\bigl(e^{-i\tau_2\omega}-1\bigr)\frac{e^{-im\omega T}-1}{e^{-i\omega T}-1},\\
        \beta_1&=-m\omega T,
    \end{aligned}
\end{equation*}
for the center-of-mass mode $b_1$ and
\begin{equation*}
    \begin{aligned}
        Z_k&=\mathrm{arcsinh}\bigl(U_{m-1}(s_k)\abs{B_k}\bigr)\frac{B_ku_k}{\abs{B_ku_k}}e^{i\phi_k},\\
        J_k&=\biggl[\sqrt{1+U_{m-1}^2(s_k)\abs{B_k}^2}e^{i\phi_k}\Bigl(\Tilde{s}_kW_m^{(k)}-W_{m-1}^{(k)}\Bigr)\\
        &\quad -\abs{B_k}^2U_{m-1}(s_k)W_m^{(k)}\frac{u_k}{\abs{u_k}}e^{i\phi_k}\biggr]G_k\\
        &\quad+B_k\biggl[U_{m-1}(s_k)\frac{u_k}{\abs{u_k}}\Bigl(\Tilde{s}_kW_m^{(k)}-W_{m-1}^{(k)}\Bigr)\\
        &\quad -\sqrt{1+U_{m-1}^2(s_k)\abs{B_k}^2}W_m^{(k)}\biggr]G_k^*,\\
        \beta_k&=\mathrm{arg}(u_k),\\
        u_k&=U_{m-1}(s_k)\Tilde{s}_k-U_{m-2}(s_k),
        \end{aligned}
\end{equation*}
\begin{equation*}
    \begin{aligned}
        W_m^{(k)}&=\sum_{l=0}^{m-1}U_l(s_k),\\
        s_k&=\mathrm{Re}(\Tilde{s}_k),\\
        \Tilde{s}_k&=A_ke^{-i\phi_k},\\
        \phi_k&=\omega T+(\Tilde{\omega}_k-\omega)\tau_2,\\
        G_k&=\frac{x_0V_1}{2\Tilde{\omega}_k}(Q_{Nk}-Q_{1k})\\
        &\quad\times \biggl(e^{-2i\tau_2\Tilde{\omega}_k}-1-\frac{\omega}{\Tilde{\omega}_k}\Bigl(e^{-i\tau_2\Tilde{\omega}_k}-1\Bigr)^2\biggr),\\
        A_k&=1+\frac{1}{2}\Bigl(1-e^{2i\tau_2\Tilde{\omega}_k}\Bigr)\frac{(\Tilde{\omega}_k-\omega)^2}{2\omega\Tilde{\omega}_k},\\
        B_k&=\Bigl(1-e^{-2i\tau_2\Tilde{\omega}_k}\Bigr)\frac{\Tilde{\omega}_k^2-\omega^2}{4\omega\Tilde{\omega}_k},\\
        \Tilde{\omega}_k&=\sqrt{\omega^2+2\lambda_kx_0^2V_2\omega},
    \end{aligned}
\end{equation*}
for all relative modes $b_k$ with $k\geq2$. We use the Chebyshev polynomials of the second kind $U_l$, which are defined by the recurrence relation
\begin{equation*}
    \begin{aligned}
        U_{-2}(x)&=-1,\\
        U_{-1}(x)&=0,\\
        U_{l}(x)&=2xU_{l-1}(x)-U_{l-2}(x)\quad\forall l\in\mathbb{N}_0.
    \end{aligned}
\end{equation*}
With these expressions, the positional covariances between the $j$-th and $k$-th atoms are given by
\begin{align}
    \mathcal{C}_{j,k}=x_0^2\sum_{l=1}^NQ_{jl}Q_{kl}\abs{\Omega_l}^2,
    \label{eq:covariances_end}
\end{align}
where we defined
\begin{align}
    \Omega_l=\sqrt{1+U_{m-1}^2(s_l)\abs{B_l}^2}-\frac{u_l}{\abs{u_l}}U_{m-1}(s_l)B_le^{i\phi_l}.\nonumber
\end{align}
The expectation values of the atomic positions take the form
\begin{align}
    \langle x_j\rangle=2x_0\sum_{l=1}^NQ_{jl}\mathrm{Re}(\Omega_l^*J_l).
    \label{eq:positions_end}
\end{align}
Using Eqs.~(\ref{eq:covariances_end}) and~(\ref{eq:positions_end}), we can determine the variances
\begin{align}
    \mathrm{Var}(\delta_{jk})=x_0^2\sum_{l=2}^N(Q_{jl}-Q_{kl})^2\abs{\Omega_l}^2
    \label{eq:var_delta}
\end{align}
and expectation values
\begin{align}
    \langle\delta_{jk}\rangle=2x_0\sum_{l=2}^N(Q_{kl}-Q_{jl})\mathrm{Re}(\Omega_l^*J_l)
    \label{eq:exp_delta}
\end{align}
of the interatomic distances $\delta_{jk}$, thereby fully characterizing their Gaussian probability distributions.


\clearpage
\onecolumngrid

\setcounter{equation}{0}
\setcounter{figure}{0}
\setcounter{table}{0}
\makeatletter
\renewcommand{\theequation}{S\arabic{equation}}
\renewcommand{\thefigure}{S\arabic{figure}}

\makeatletter
\renewcommand{\theequation}{S\arabic{equation}}
\renewcommand{\thefigure}{S\arabic{figure}}

\begin{center}
{\Large SUPPLEMENTAL MATERIAL}
\end{center}
\begin{center}
\vspace{0.8cm}
{\Large Creating squeezed and non-classical collective motional many-body states through stroboscopic Rydberg dressing}
\end{center}
\begin{center}
Roman Wu\ss ler$^{1}$, Chris Nill$^{1,2}$, Sylvain de L\'es\'eleuc$^{3,4}$, Christian Gro\ss$^{5}$, Igor Lesanovsky$^{1,6}$
\end{center}
\begin{center}
$^1${\em Institut f\"ur Theoretische Physik, Universit\"at T\"ubingen,}
{\em Auf der Morgenstelle 14, 72076 T\"ubingen, Germany}\\
$^2${\em Institute for Applied Physics, University of Bonn, Wegelerstraße 8, 53115 Bonn, Germany}\\
$^3${\em Institute for Molecular Science, National Institutes of Natural Sciences, 444-8585 Okazaki, Japan}\\
$^4${\em RIKEN Center for Quantum Computing (RQC), 351-0198 Wako, Japan}\\
$^5${\em Physikalisches Institut and Center for Integrated Quantum Science and Technology, Universit\"{a}t T\"{u}bingen, Auf der Morgenstelle 14, 72076 T\"{u}bingen, Germany}\\
$^6${\em School of Physics and Astronomy and Centre for the Mathematics and Theoretical Physics of Quantum Non-Equilibrium Systems, The University of Nottingham, Nottingham, NG7 2RD, United Kingdom}
\end{center}

\section{Derivation of the time evolution operator \texorpdfstring{$\mathcal{U}(mT)$}{U(mT)} at stroboscopic times}
The following derivation yields an intuitive, closed-form expression for the time evolution operator $\mathcal{U}(mT)$. It describes the evolution of the system during $m$ dressing cycles of the stroboscopic dressing protocol, which is illustrated in Figs.~\ref{fig:cartoon}(b) and~(c) in the main text. Each dressing cycle is decomposed into four phases. During the first one of duration $\tau_1$, the system evolves freely under the Rydberg Hamiltonian ($\hbar=1$)
\begin{align}
    H_0=\omega\sum_{j=1}^Na_j^\dagger a_j+\sum_{j=1}^{N-1}n_jn_{j+1}\biggl(V_0+V_1(x_{j+1}-x_j)+\frac{V_2}{2}(x_{j+1}-x_j)^2\biggr),
    \label{eq:H_0_sm}
\end{align}
where $\omega$ is the trapping frequency of the optical tweezers with the ladder operators $a_j$ and $a_j^\dagger$, describing the quantized motion of the atoms along the chain axis. The Rydberg interaction potential is assumed to be a van der Waals potential $V(r)=C_6/r^6$, where $r$ is the interatomic distance of two atoms. Employing a second-order Taylor expansion leads to the Rydberg interaction term in Eq.~\ref{eq:H_0_sm}. The potential is parameterized by its interaction strength $V_0=C_6/a_0^6$, gradient $V_1=-6C_6a_0^7$, and curvature $V_2=42C_6/a_0^8$, evaluated at the equilibrium distance $a_0$ between neighboring atoms. Since $n_j=\ketbra{\uparrow}{\uparrow}_j$ is the projector onto the Rydberg state of the $j$-th atom, this term only contributes if two neighboring atoms are excited to the Rydberg state. The operator
\begin{align}
    x_j=x_0\Bigl(a_j+a_j^\dagger\Bigr)=\frac{1}{\sqrt{2m_0\omega}}\Bigl(a_j+a_j^\dagger\Bigr)
\end{align}
describes the displacement of the $j$-th atom out of its equilibrium position, where $m_0$ denotes the atomic mass.\\
During the second phase of duration $s$, the system is driven by a laser with Rabi frequency $\Omega=\pi/s$, corresponding to a resonant $\pi$-pulse, which couples the ground state $\ket{\downarrow}$ to the Rydberg state $\ket{\uparrow}$. This leads to the additional laser Hamiltonian
\begin{align}
    H_\mathrm{L}=\frac{\pi}{2s}\sum_{j=1}^N\Bigl(\ketbra{\downarrow}{\uparrow}_je^{-i\kappa x_j}+\ketbra{\uparrow}{\downarrow}_je^{i\kappa x_j}\Bigr),
\end{align}
where $\kappa$ is the projection of the wave vector onto the chain axis. We consider the limit of strong laser pulses, such that $\pi/s=\Omega\gg\omega, V_0, x_0V_1, x_0^2V_2$ is by orders of magnitude the largest energy scale in the system, and pulses can break the Rydberg blockade. Thus, during the laser pulses, we can neglect the Rydberg Hamiltonian $H_0$ and only consider the evolution of the system under the laser Hamiltonian $H_\mathrm{L}$, see End Matter.
After the first $\pi$-pulse was applied, the laser is switched off again. The system then evolves freely under the Rydberg Hamiltonian $H_0$ for a time $\tau_2$, before a second $\pi$-pulse of duration $s$ is applied, completing the dressing cycle. Consequently, the time evolution operator for one dressing cycle reads
\begin{align}
    \mathcal{U}(T)=\underbrace{e^{-isH_\mathrm{L}}}_{\mathcal{U}_\pi}e^{-i\tau_2H_0}e^{-isH_\mathrm{L}}e^{-i\tau_1H_0}.
\end{align}
The time evolution operator of a single $\pi$-pulse can be simplified to
\begin{align}
    \mathcal{U}_\pi=\exp\left\{-i\frac{\pi}{2}\sum_{j=1}^N\Bigl(\ketbra{\downarrow}{\uparrow}_je^{-i\kappa x_j}+\ketbra{\uparrow}{\downarrow}_je^{i\kappa x_j}\Bigr)\right\}=(-i)^N\prod_{j=1}^N\Bigl(\ketbra{\downarrow}{\uparrow}_je^{-i\kappa x_j}+\ketbra{\uparrow}{\downarrow}_je^{i\kappa x_j}\Bigr).
\end{align}
By utilizing that $\mathcal{U}_\pi$ is unitary, the time evolution operator $\mathcal{U}(T)$ can be written as
\begin{align}
    \mathcal{U}(T)=\mathcal{U}_\pi\underbrace{\mathcal{U}_\pi\mathcal{U}_\pi^\dagger}_{=1}e^{-i\tau_2H_0}\mathcal{U}_\pi e^{-i\tau_1H_0}=\underbrace{\mathcal{U}_\pi^2}_{=(-1)^N}e^{-i\tau_2\mathcal{U}_\pi^\dagger H_0 U_\pi}e^{-i\tau_1H_0}.
\end{align}
We define the Lamb-Dicke parameter $\eta=\kappa x_0$ and the projector $\mathcal{P}_j=\ketbra{\downarrow}{\downarrow}_j$ onto the ground state $\ket{\downarrow}_j$ of the $j$-th atom. The relation $\mathcal{U}_\pi^\dagger n_j\mathcal{U}_\pi=\mathcal{P}_j$ shows that the stroboscopic dressing protocol maps the interactions between atoms in the Rydberg states to effective interactions between ground-state atoms. Using this identity together with $\mathcal{U}_\pi^\dagger a_j\mathcal{U}_\pi=a_j-i\eta(n_j-\mathcal{P}_j)$, we obtain
\begin{equation}
    \begin{aligned}
        \mathcal{U}(T)&=(-1)^N\exp\left\{-i\tau_2\left[\omega\sum_{j=1}^N\Bigl(a_j^\dagger a_j+i\eta(n_j-\mathcal{P}_j)\Bigl(a_j-a_j^\dagger\Bigr)+\eta^2\Bigr)\right.\right.\\
        &\quad\left.\left.+\sum_{j=1}^{N-1}\mathcal{P}_j\mathcal{P}_{j+1}\biggl(V_0+V_1(x_{j+1}-x_j)+\frac{V_2}{2}(x_{j+1}-x_j)^2\biggr)\right]\right\}\\
        &\quad\cdot\exp\left\{-i\tau_1\left[\omega\sum_{j=1}^Na_j^\dagger a_j+\sum_{j=1}^{N-1}n_jn_{j+1}\biggl(V_0+V_1(x_{j+1}-x_j)+\frac{V_2}{2}(x_{j+1}-x_j)^2\biggr)\right]\right\}.
    \end{aligned}
\end{equation}
All atoms are initialized in the electronic ground state, i.e. $\ket{\Psi(0)}=\ket{\downarrow}^{\otimes N}\otimes\ket{\Psi_\mathrm{motion}(0)}$. Since the operators in the exponents of $\mathcal{U}(T)$ are diagonal in the electronic basis, the system never leaves the electronic ground state at stroboscopic times. Therefore, we can project the dynamics onto this subspace by replacing the projectors with their respective eigenvalues, i.e. $n_j\to0$ and $\mathcal{P}_j\to1$. This leads to
\begin{equation}
    \begin{aligned}
        \mathcal{U}(T)&=(-1)^Ne^{-i\tau_2(N\omega\eta^2+(N-1)V_0)}\exp\left\{-i\tau_2\left[\omega\sum_{j=1}^N\Bigl(a_j^\dagger a_j-i\eta\Bigl(a_j-a_j^\dagger\Bigr)\Bigr)\right.\right.\\
        &\quad\left.\left.+\sum_{j=1}^{N-1}\biggl(V_1(x_{j+1}-x_j)+\frac{V_2}{2}(x_{j+1}-x_j)^2\biggr)\right]\right\}\exp\left\{-i\tau_1\omega\sum_{j=1}^Na_j^\dagger a_j\right\}.
    \end{aligned}
\end{equation}
Due to the finite curvature of the Rydberg interaction potential, the time evolution operator contains mixed-mode terms $x_{j+1}x_j$ that couple adjacent sites. To diagonalize the quadratic term of the exponential and further simplify $\mathcal{U}(T)$, we introduce the normal modes of the system, given by
\begin{align}
    b_j=\sum_{k=1}^NQ_{kj}a_k,
\end{align}
with the transformation matrix elements
\begin{align}
    Q_{kj}=\begin{cases}
        \frac{1}{\sqrt{N}} & \text{for } j=1,\\
        \sqrt{\frac{2}{N}}\cos\biggl(\frac{(j-1)\bigl(k-\frac{1}{2}\bigr)\pi}{N}\biggr) & \text{for } 2\le j\le N.
    \end{cases}
\end{align}
Since the transformation matrix $Q$ is orthogonal, the new operators satisfy the bosonic commutation relations $[b_j, b_k^\dagger]=\delta_{jk}$ and $[b_j,b_k]=[b_j^\dagger, b_k^\dagger]=0$. The mode $b_1=\sum_{k=1}^Na_k/\sqrt{N}$ is the center-of-mass mode of the system, while all other modes $b_j$ with $j>1$ describe the relative motion of the atoms. Inserting the normal modes into the time evolution operator $\mathcal{U}(T)$ for a single dressing cycle, and raising it to the power of $m$, we obtain the time evolution operator for $m$ dressing cycles, given by
\begin{equation}
    \begin{aligned}
        \mathcal{U}(mT)&=(-1)^{mN}e^{-im\tau_2\left(N\omega\eta^2+(N-1)\left(V_0+x_0^2V_2\right)\right)}\Bigl(\exp\Bigl\{-i\tau_2\omega\Bigl[b_1^\dagger b_1-i\sqrt{N}\eta\Bigl(b_1-b_1^\dagger\Bigr)\Bigr]\Bigr\}\exp\Bigl\{-i\tau_1\omega b_1^\dagger b_1\Bigr\}\Bigr)^m\\
        &\quad\cdot\prod_{k=2}^N\biggl(\exp\biggl\{-i\tau_2\biggl[\bigl(\omega+\lambda_kx_0^2V_2\bigr)b_k^\dagger b_k+x_0V_1(Q_{Nk}-Q_{1k})\Bigl(b_k^\dagger+b_k\Bigr)+\frac{\lambda_kx_0^2V_2}{2}\biggl(b_k^{\dagger2}+b_k^2\biggr)\biggr]\biggr\}\\
        &\quad\cdot\exp\Bigl\{-i\tau_1b_k^\dagger b_k\Bigr\}\biggr)^m\\
        &=\prod_{k=1}^N\mathcal{U}_k(mT),
    \end{aligned}
\end{equation}
where we defined
\begin{align}
    \lambda_k=4\sin^2\biggl(\frac{(k-1)\pi}{2N}\biggr).
\end{align}
The time evolution operator factorizes into a product over all $N$ normal modes of the system, indicating that they decouple and evolve independently. Since the operators $\mathcal{U}_k(mT)$ are Gaussian, meaning that they contain at most quadratic bosonic operators in their exponents, we can proceed by applying a Bloch-Messiah decomposition~\cite{cariolaroReexaminationBlochMessiahReduction2016} to each mode separately. Specifically, this allows us to express the time evolution operator as an ordered sequence of fundamental Gaussian unitaries
\begin{align}
    \mathcal{U}_k(mT)=\mathcal{S}_k(Z_k)\mathcal{D}_k(J_k)\mathcal{R}_k(\beta_k)e^{i\gamma_k},
\end{align}
where $\mathcal{S}_k(Z_k)=e^{\frac{Z_k^*}{2}b_k^2-\frac{Z_k}{2}b_k^{\dagger2}}$, $\mathcal{D}_k(J_k)=e^{J_kb_k^\dagger-J_k^*b_k}$ and $\mathcal{R}_k(\beta_k)=e^{i\beta_kb_k^\dagger b_k}$ denote the squeezing, displacement and rotation operators respectively. To calculate the parameters $Z_k$, $J_k$, $\beta_k$, and $\gamma_k$ for each mode, we exploit the algebraic structure determined by the operators $\{1, b_k, b_k^\dagger, b_k^2, b_k^{\dagger2}, b_k^\dagger b_k\}$. Therefore, we define the six matrices
\begin{equation}
    \begin{aligned}
        M_1&=\begin{pmatrix}
                0 & 0 & 0 & 1 \\
                0 & 0 & 0 & 0 \\
                0 & 0 & 0 & 0 \\
                0 & 0 & 0 & 0
            \end{pmatrix},\quad
        M_b=\frac{1}{\sqrt{2}}\begin{pmatrix}
                0 & 0 & -1 & 0 \\
                0 & 0 & 0 & 1 \\
                0 & 0 & 0 & 0 \\
                0 & 0 & 0 & 0
            \end{pmatrix},\quad
        M_{b^\dagger}=\frac{1}{\sqrt{2}}\begin{pmatrix}
                0 & -1 & 0 & 0 \\
                0 & 0 & 0 & 0 \\
                0 & 0 & 0 & -1 \\
                0 & 0 & 0 & 0
            \end{pmatrix},\\
        M_{b^2}&=\begin{pmatrix}
                0 & 0 & 0 & 0 \\
                0 & 0 & -2 & 0 \\
                0 & 0 & 0 & 0 \\
                0 & 0 & 0 & 0
            \end{pmatrix},\quad
        M_{b^{\dagger2}}=\begin{pmatrix}
                0 & 0 & 0 & 0 \\
                0 & 0 & 0 & 0 \\
                0 & 2 & 0 & 0 \\
                0 & 0 & 0 & 0
            \end{pmatrix},\quad
        M_{b^\dagger b}=\begin{pmatrix}
                0 & 0 & 0 & -\frac{1}{2} \\
                0 & -1 & 0 & 0 \\
                0 & 0 & 1 & 0 \\
                0 & 0 & 0 & 0
            \end{pmatrix},
    \end{aligned}
\end{equation}
which satisfy the same commutation relations as the corresponding bosonic operators, forming a faithful finite-dimensional matrix representation of the algebra. The Bloch-Messiah decomposition can be performed using the Baker-Campbell-Hausdorff (BCH) formula, which relies exclusively on the commutation relations among the involved operators. Therefore, the decomposition is entirely governed by the structure of the underlying Lie algebra. Since the matrices and the corresponding bosonic operators are two representations of the same Lie algebra, the time evolution matrix can be decomposed in the same way as the original bosonic operator. Consequently, we can replace the operators with their corresponding matrices, compute the matrix exponentials, and equate the result to the target matrix representing the operator $\mathcal{S}_k(Z_k)\mathcal{D}_k(J_k)\mathcal{R}_k(\beta_k)e^{i\gamma_k}$. By solving the resulting system of equations, we can calculate all parameters of interest. Up to a global phase that does not affect any expectation values, the time evolution operator is given by
\begin{align}
    \mathcal{U}(mT)=\prod_{k=1}^N\mathcal{S}_k(Z_k)\mathcal{D}_k(J_k)\mathcal{R}_k(\beta_k),
\end{align}
with the parameters
\begin{equation}
    \begin{aligned}
        Z_1&=0,\\
        J_1&=\sqrt{N}i\eta\bigl(e^{-i\tau_2\omega}-1\bigr)\frac{e^{-im\omega T}-1}{e^{-i\omega T}-1},\\
        \beta_1&=-m\omega T,
    \end{aligned}
\end{equation}
for the center-of-mass mode $b_1$ and
\begin{equation}
    \begin{aligned}
        Z_k&=\mathrm{arcsinh}\bigl(U_{m-1}(s_k)\abs{B_k}\bigr)\frac{B_ku_k}{\abs{B_ku_k}}e^{i\phi_k},\\
        J_k&=\biggl[\sqrt{1+U_{m-1}^2(s_k)\abs{B_k}^2}e^{i\phi_k}\Bigl(\Tilde{s}_kW_m^{(k)}-W_{m-1}^{(k)}\Bigr)-\abs{B_k}^2U_{m-1}(s_k)W_m^{(k)}\frac{u_k}{\abs{u_k}}e^{i\phi_k}\biggr]G_k\\
        &\quad+B_k\biggl[U_{m-1}(s_k)\frac{u_k}{\abs{u_k}}\Bigl(\Tilde{s}_kW_m^{(k)}-W_{m-1}^{(k)}\Bigr)-\sqrt{1+U_{m-1}^2(s_k)\abs{B_k}^2}W_m^{(k)}\biggr]G_k^*,\\
        \beta_k&=\mathrm{arg}(u_k),\\   
        u_k&=U_{m-1}(s_k)\Tilde{s}_k-U_{m-2}(s_k),\\
        W_m^{(k)}&=\sum_{l=0}^{m-1}U_l(s_k),\\
        s_k&=\mathrm{Re}(\Tilde{s}_k),\\
        \Tilde{s}_k&=A_ke^{-i\phi_k},\\
        \phi_k&=\omega T+(\Tilde{\omega}_k-\omega)\tau_2,\\
        G_k&=\frac{x_0V_1}{2\Tilde{\omega}_k}(Q_{Nk}-Q_{1k})\biggl(e^{-2i\tau_2\Tilde{\omega}_k}-1-\frac{\omega}{\Tilde{\omega}_k}\Bigl(e^{-i\tau_2\Tilde{\omega}_k}-1\Bigr)^2\biggr),\\
        A_k&=1+\frac{1}{2}\Bigl(1-e^{2i\tau_2\Tilde{\omega}_k}\Bigr)\frac{(\Tilde{\omega}_k-\omega)^2}{2\omega\Tilde{\omega}_k},\\
        B_k&=\Bigl(1-e^{-2i\tau_2\Tilde{\omega}_k}\Bigr)\frac{\Tilde{\omega}_k^2-\omega^2}{4\omega\Tilde{\omega}_k},\\
        \Tilde{\omega}_k&=\sqrt{\omega^2+2\lambda_kx_0^2V_2\omega},
    \end{aligned}
\end{equation}
for all relative modes $b_k$ where $k$ is a natural number between $2$ and $N$. We used the Chebyshev polynomials of the second kind $U_l$, which are defined by the recurrence relation
\begin{equation}
    \begin{aligned}
        U_{-2}(x)&=-1,\\
        U_{-1}(x)&=0,\\
        U_{l}(x)&=2xU_{l-1}(x)-U_{l-2}(x)\quad\forall l\in\mathbb{N}_0.
    \end{aligned}
\end{equation}

\section{Variances, covariances and expectation values of the interatomic distances}
To quantify the motional squeezing discussed in the main text, we derive in this section the explicit analytical expressions for the variances, covariances, and expectation values of the interatomic distances. We assume the system to be initialized in the electronic and motional ground state $\ket{\Psi(0)}=\ket{\downarrow,0}^{\otimes N}$. Consequently, after $m$ dressing cycles, the system (up to a global phase that does not affect any expectation values) evolves to
\begin{align}
    \ket{\Psi(mT)}=\mathcal{U}(mT)\ket{\Psi(0)}=\ket{\downarrow}^{\otimes N}\prod_{k=1}^N\mathcal{S}_k(Z_k)\mathcal{D}_k(J_k)\ket{0}.
\end{align}
By utilizing the transformations $\mathcal{S}_k^\dagger(r_ke^{i\theta_k})b_k\mathcal{S}_k(r_ke^{i\theta_k})=\cosh(r_k)b_k-\sinh(r_k)e^{i\theta_k}b_k^\dagger$ and $\mathcal{D}_k^\dagger(J_k)b_k\mathcal{D}_k(J_k)=b_k+J_k$ of the annihilation operator $b_k$ under the squeezing and displacement operators, we can calculate the expectation values of arbitrary bosonic operators. The positional covariances between two atoms are given by
\begin{align}
    \mathcal{C}_{j,k}=\langle x_jx_k\rangle-\langle x_j\rangle\langle x_k\rangle=x_0^2\sum_{l=1}^NQ_{jl}Q_{kl}\abs{\Omega_l}^2,
    \label{eq:covariances}
\end{align}
where we defined
\begin{align}
    \Omega_l=\sqrt{1+U_{m-1}^2(s_l)\abs{B_l}^2}-\frac{u_l}{\abs{u_l}}U_{m-1}(s_l)B_le^{i\phi_l}.
\end{align}
The expectation values of the atomic positions are given by
\begin{align}
    \langle x_j\rangle=2x_0\sum_{l=1}^NQ_{jl}\mathrm{Re}(\Omega_l^*J_l).
    \label{eq:positions}
\end{align}
Using Eqs.~\ref{eq:covariances} and~\ref{eq:positions}, we can determine the variances
\begin{align}
    \mathrm{Var}(\delta_{jk})=\underbrace{\mathrm{Var}(x_j)}_{=\mathcal{C}_{j,j}}+\mathrm{Var}(x_k)-2\mathcal{C}_{j,k}=x_0^2\sum_{l=2}^N(Q_{jl}-Q_{kl})^2\abs{\Omega_l}^2
\end{align}
and expectation values
\begin{align}
    \langle\delta_{jk}\rangle=\langle x_k\rangle-\langle x_j\rangle=2x_0\sum_{l=2}^N(Q_{kl}-Q_{jl})\mathrm{Re}(\Omega_l^*J_l)
\end{align}
of the interatomic distances $\delta_{jk}=x_k-x_j$, thereby fully characterizing their Gaussian probability distributions. These analytical expressions form the basis of the motional dynamics of the system, presented in Fig.~\ref{fig:mean_and_covariances} of the main text.

\section{Periodicities and protocol stability}
For the stroboscopic dressing protocol to remain stable over many dressing cycles, the atoms must stay within their trapping potentials, requiring their individual mean displacements $\langle x_j\rangle$ to be tightly bound. To understand the conditions for this stability, shown in Fig.~\ref{fig:mean_and_covariances}~(e)~and~(f), we need to analyze the periodic behavior of the variances of the interatomic distances and the expectation values of the individual atomic displacements with respect to the phase $\omega T$. We will show in the following that while the variances are $\pi$-periodic in $\omega T$, the mean displacements are strictly $2\pi$-periodic.\\
To demonstrate this, we focus on the phase factor $e^{-i\phi_l}=e^{-i(\omega T+(\Tilde{\omega}_l-\omega)\tau_2)}$, as it contains the sole dependence on the phase $\omega T$ in our analytical expressions. Under a shift of $\omega T\to\omega T+\pi$, it acquires a minus sign, $e^{-i\phi_l}\to-e^{-i\phi_l}$, which implies $\Tilde{s}_l\to-\Tilde{s}_l$ and, for its real part, $s_l\to-s_l$. Using the parity relation $U_m(-x)=(-1)^mU_m(x)$ of the Chebyshev polynomials, it follows that $u_l=U_{m-1}(s_l)\Tilde{s}_l-U_{m-2}(s_l)$ transforms as $u_l\to(-1)^mu_l$. Therefore, $\Omega_l\to\Omega_l$ remains invariant, implying that $\mathcal{C}_{j,k}$ and $\mathrm{Var}(\delta_{jk})$ also remain invariant. Thus, they are $\pi$-periodic in $\omega T$.\\
On the other hand, the expectation values $\langle x_{j}\rangle$ depend linearly on the displacement $J_l$, which includes the sum $W_m^{(k)}=\sum_{l=0}^{m-1}U_l(s_k)$. Since this sum mixes Chebyshev polynomials of both, even and odd degrees, it does not possess a simple global parity under $\omega T\to\omega T+\pi$. As a result, $J_l$ does not return to its initial value under a $\pi$ shift. A full shift of $2\pi$ is required to restore $J_l\to J_l$, demonstrating that the expectation values $\langle x_j\rangle$ are strictly $2\pi$-periodic in $\omega T$.\\
Knowing these periodicities allows us to optimize the cycle duration $T$, such that the variances $\mathrm{Var}(\delta_{jk})$ of the interatomic distances become minimal while the individual atoms stay tightly bound in their traps. Since $\mathrm{Var}(\delta_{jk})$ is $\pi$-periodic in $\omega T$, we first determine $T_{\mathrm{opt}}$ by minimizing these variances within the half-period $\omega T\in[0,\pi)$. Because the mean displacements $\langle x_j\rangle$ are strictly $2\pi$-periodic, the choices $\omega T_{\mathrm{opt}}$ and $\omega T_{\mathrm{opt}}+\pi$ yield the same minimized variances but result in different atomic displacements. To prevent atom loss, we evaluate both regimes and select the cycle duration that minimizes the mean displacements $\langle x_j\rangle$, as seen in Fig.~\ref{fig:mean_and_covariances}~(e) and~(f).

\section{Anharmonic tweezer potential}
To quantify the impact of trap anharmonicities, we extend the harmonic potential by a quartic perturbation, following~\cite{Lienhard2025}. The corresponding correction to the Hamiltonian $H_0$ reads
\begin{equation}
    H_{\text{anharm}} = -\frac{\omega \epsilon}{4} \sum_{j=1}^N (a_j + a_j^\dagger)^4,
    \label{eq:H_anharmonic}
\end{equation}
where the parameter $\epsilon$ quantifies the anharmonicity of the trapping potential. 

\begin{figure}[t]
    \centering
    \includegraphics[width=.95\linewidth]{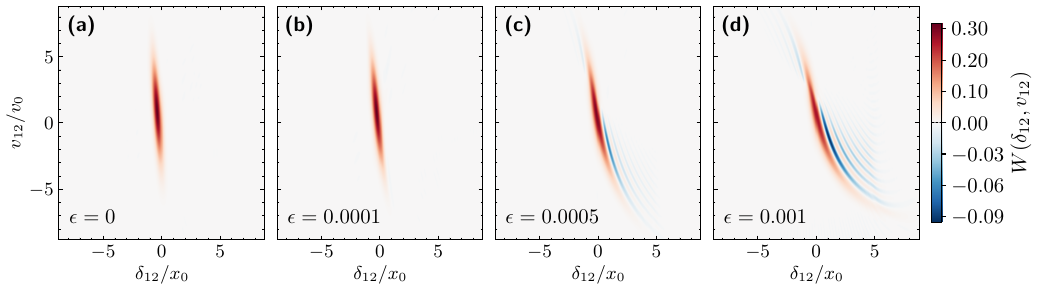}
    \caption{\textbf{Impact of anharmonic tweezer potentials on motional squeezing.} Wigner functions $W(\delta_{12}, v_{12})$ of the relative coordinate $\delta_{12}$ and relative velocity $v_{12}$ for $N=2$ atoms after $m=20$ dressing cycles, computed including the anharmonic trap Hamiltonian from Eq.~\eqref{eq:H_anharmonic}. \textbf{(a)} For a perfectly harmonic trap ($\epsilon=0$), the motional state is Gaussian and well-described by the analytical prediction from Eq.~(\ref{eq:covariances}). \textbf{(b)--(d)} With increasing anharmonicity $\epsilon$, the Wigner function experiences progressive shearing and non-Gaussian deformations, leading to the emergence of negative regions (indicated in blue). All panels use the parameters from Eq.~(\ref{eq:parameter-set-1}).}
    \label{fig:anharmonic_wigners}
\end{figure}

In Fig.~\ref{fig:anharmonic_wigners}, we vary $\epsilon$ to evaluate the impact of anharmonic tweezer potentials on the motional state. For sufficiently harmonic traps with $\epsilon\leq10^{-4}$, the motional state remains Gaussian (a),~(b) and can be described by our analytical model from Eq.~\eqref{eq:covariances}. However, for increased values of $\epsilon$ (c),~(d), the Wigner function experiences significant shearing, and pronounced negative regions emerge~\cite{Rosiek2024}. Current standard optical tweezers typically exhibit anharmonicities on the order of $\epsilon\approx10^{-3}$~\cite{Lienhard2025}.
While realizing purely Gaussian, strongly squeezed states will require improved trap harmonicity, this inherent anharmonicity is not strictly a limitation. Instead, it can be actively leveraged as a resource for generating genuine non-classical many-body motional states~\cite{Grochowski2025}.



\end{document}